\documentclass{article}
\usepackage{amsmath,amssymb}
\usepackage[left=1in,right=1in,top=1in,bottom=1in]{geometry}
\usepackage{graphicx}
\usepackage{authblk}
\usepackage{fourier}
\usepackage{color}
\usepackage[colorlinks=true,citecolor=blue]{hyperref}
\numberwithin{equation}{section}

\title{On the Riemann problem for the Adlam-Allen model}
\author[1]{Su Yang\thanks{Corresponding author: suyang@umass.edu}}

\author[1]{Marco Calabrese}
\author[2]{Vassilis Koukouloyannis}
\author[1,3,4]{Panayotis G.~Kevrekidis}
\affil[1]{Department of Mathematics and Statistics, University of Massachusetts Amherst, Amherst, MA 01003-4515, USA}
\affil[2]{Department of Mathematics, University of the Aegean, Karlovasi, 83200 Samos, Greece}
\affil[3]{Department of Physics, University of Massachusetts Amherst,  Massachusetts 01003}
\affil[4]{Department of Mechanical Engineering, Seoul National University, 1 Gwanak-ro, Gwanak-gu, Seoul 08826, South Korea}

\date{\small\today}

\begin{document}

\maketitle

\begin{abstract}
    In the present work, we revisit the Adlam-Allen (AA) model in
    order to investigate its numerically observed rarefaction and dispersive shock waves that arise in numerical simulations of the Riemann problem associated with the model. On the one hand, we perform a direct analysis of the rarefaction and dispersive shock waves of the AA model via examining its corresponding dispersionless system and leveraging the DSW-fitting method to obtain theoretical predictions on various edge features of the dispersive shock waves. On the other hand, we review the KdV reduction of the AA model and utilize the KdV dispersive shock wave to approximate that of the AA model. Relevant numerical comparisons demonstrate the good performance of not only the direct analysis on the AA dispersive shock wave, but also of the approximation via the KdV DSW. 
    These methodologies provide a systematic toolbox for analyzing the outcome
    of Riemann problems in not only this fundamental setting of cold
    plasmas but also potentially in related plasma-physics problems.
\end{abstract}

\section{Introduction}
The study of shock waves in dispersive systems, i.e., so-called dispersive shock waves (DSWs), has seen many developments in the last few decades. In particular, these phenomena have been extensively studied across fluid dynamics (including superfluids)~\cite{PhysRevA.74.023623,GONG2022133398}, nonlinear optics \cite{1e60d1eb29b441e3adaee681206ceb2b,PhysRevLett.118.254101}, discrete models~\cite{herbold,molinari,talcohen}, 
and plasma physics \cite{doi:10.1126/sciadv.aau9926,Woods_1972}.  A comprehensive theoretical framework for their description is provided by Whitham modulation theory \cite{whitham2011linear,EL201611} and 
further details can be systematically understood for the DSW
leading and trailing edges, e.g., via the so-called DSW fitting method~\cite{EL201611}.

The realm of plasma physics, more concretely, has been a focal point for
the analysis of the features of nonlinear waves. Indeed, this was already noted
in the 1965 landmark paper of Kruskal and Zabusky~\cite{kruskal} who referred to
unpublished work of Gardner and Morikawa (from 1960, in the form of
an unpublished Courant report) on
collisionless--plasma magnetohydrodynamic waves as having incited their interest
in the Korteweg-de Vries (KdV) equation.
Remarkably, it turns out that seminal work by Adlam and Allen
already in 1958~\cite{Adlam01051958} (see also~\cite{JHAdlam_1960}) 
had developed a modeling framework for hydromagnetic waves in cold plasmas,
in the form of what is today known as the Adlam-Allen (AA) model. 
The relevant system of nonlinear partial differential equations
---for the (reciprocal) density of carriers and the magnetic field within the
plasma--- has recently attracted some renewed attention,
both from one of the original authors~\cite{Abbas2020ASO,Abbas_2022},
as well as from some of the present authors~\cite{PhysRevE.102.013209,PhysRevE.106.034209}. This
is, at least in part, due to its rich, multi-component nonlinear structure 
leading to intriguing features (and connections with the KdV model) 
for the solitary waves~\cite{PhysRevE.102.013209}, periodic solutions~\cite{PhysRevE.106.034209}, as well as for the 
shock waves~\cite{AAshock} featured therein. 
The AA model has also been recently argued to be relevant in physical settings such as magnetospheric plasmas \cite{Abbas2020ASO,Abbas_2022,Abbas2020PropagationOP}, adding further 
physical appeal to its 
complex mathematical structure.

In its non-dissipative, dimensionless form, the AA model can be written as a coupled system  \cite{PhysRevE.102.013209} for two fields $u(x,t)$ and $w(x,t)$. Considering the physical quantities of the (rescaled) inverse
carrier density of ions and electrons $R=R_0 + u(x,t)$ and
the magnetic field $B=B_0 + w(x,t)$ (i.e., in both cases, and factoring
out the asymptotic form of these quantities for the plasma as
$x \rightarrow \pm \infty$), we obtain a nonlinear system of
partial differential equations. 
This is given by:
\begin{equation}\label{eq: AA model}
    \begin{aligned}
    &u_{tt} + B_0w_{xx} + \frac{1}{2}\left(w^2\right)_{xx} = 0,\\
    &w_{xx} - R_0w - B_0u - uw = 0,
    \end{aligned}
\end{equation}
with its mathematical form coupling a nonlinear wave equation to an elliptic constraint.
Throughout this work, we  set $R_0 = B_0 = 1$~\footnote{In much of
what follows, we will still use the symbolism of
$R_0$ and $B_0$ to present the results as generally as 
possible. Yet, our explicit numerical computations
will always use $R_0 = B_0 = 1$}.
As indicated above, the AA model $(\ref{eq: AA model})$ supports both solitary and periodic traveling waves \cite{PhysRevE.102.013209}. It is 
well-known~\cite{whitham2011linear,EL201611} that the latter form the key building block for the emergence of DSWs. Indeed, these can be interpreted as slowly modulated periodic solutions 
(i.e., with their amplitude, wavenumber, and wave mean being slowly 
varying), as described ---to leading order---
by Whitham modulation theory.

Another set of insights into the AA model is obtained through asymptotic reduction. Indeed, in the small-amplitude, long wave regime, the system of Eq.~$(\ref{eq: AA model})$ reduces to the Korteweg-de Vries (KdV) equation as shown in~\cite{NAIRN_BINGHAM_ALLEN_2005} (see also~\cite{PhysRevE.102.013209}), establishing a direct link with classical DSW theory and enabling explicit predictions for amplitudes and propagation speeds.
Motivated by these developments, in this work we systematically investigate dispersive shock waves in the Adlam-Allen model using the tools mentioned above,
namely Whitham modulation theory, DSW fitting, as well as asymptotic reductions and leveraging of the well-established KdV DSW theory; these are all complemented by and corroborated through direct numerical simulations. 
We thus provide a systematic characterization of the formation and structure of AA DSWs arising from Riemann-type initial data.

Our presentation is structured as follows. In section 2 we discuss the
solitary and periodic traveling wave solutions to the AA model.
In section 3 we derive the Whitham modulation theory for the model.
In section 4 we analyze the Riemann problem, while in section 5
we provide the analytical considerations of the DSW fitting. 
The rarefaction wave solution is briefly discussed in section 6,
while section 7 provides a numerical validation of the different
theoretical concepts that we present. In section 8 we provide
the link to the KdV DSW analysis. Finally, in section 9, we summarize
our conclusions, as well as present some directions for future studies.

\section{Traveling wave solutions}

\subsection{Solitary Waves}
In this section, we review the traveling solitary-wave solution of the AA model. The solitary-wave solution is important in the sense that it  
arises at the leading edge of the AA DSW which is the main object  investigated throughout this work.
In the description that follows, we use the notation 
of~\cite{PhysRevE.102.013209}, according to which the traveling solitary-wave solution to the AA model \eqref{eq: AA model} reads
\begin{equation}\label{e: soliton soln}
   \begin{aligned}
    w(x,t) &= \frac{2B_0}{\gamma}\left(c^2-\gamma^2\right)\frac{1}{\gamma+c\cosh\left(\theta\right)},\\
    \theta &= \frac{B_0}{c\gamma}\sqrt{c^2-\gamma^2}\left(x-ct-x_0\right),
    \end{aligned}
\end{equation}
where $\gamma^2 = B_0^2/R_0$ and $x_0$ denotes the arbitrary phase parameter.

Then, it is straightforward to see that the amplitude of the solitary wave, denoted as $a$, in Eq.~\eqref{e: soliton soln} reads
\begin{equation}\label{e: soliton amplitude}
    a = \frac{2B_0(c-\gamma)}{\gamma},
\end{equation}
which specifies the soliton ``amplitude-speed" relation. 
Moreover, the (inverse) density-related field $u(x,t)$ can
be directly inferred using the second equation of~(\ref{eq: AA model}).

\subsection{Periodic traveling wave solutions}\label{Subsec: periodic solutions}

Besides the traveling solitary-wave solution, it is also necessary to compute periodic traveling-wave solution to the AA model \eqref{eq: AA model}, since the AA DSW essentially stems from the  periodic solutions
and, in particular, their slow, self-similar modulation.

To compute the periodic solutions of the AA model \eqref{eq: AA model}, we assume the following traveling-wave ansatz:
\begin{equation}\label{eq: Traveling-wave ansatz}
    u(x,t) = U(z), \quad w(x,t) = W(z); \quad z = x - ct,
\end{equation}
where $c \in \mathbb{R}$ refers to the propagation speed  of the traveling wave.

We substitute the ansatz \eqref{eq: Traveling-wave ansatz} into the system \eqref{eq: AA model} and obtain
\begin{equation}\label{eq: substitution of traveling-wave ansatz}
   \begin{aligned}
    &U = \frac{1}{c^2}\left(E-\frac{1}{2}W^2-B_0W\right),\\
    &W_{zz} - R_0W - B_0U - UW = 0,
    \end{aligned}
\end{equation}
where $E$ is a constant of integration.

Then, we substitute the first equation in system Eq.~\eqref{eq: substitution of traveling-wave ansatz} into the second one to obtain that,
\begin{equation}\label{eq: second sub}
    W_{zz} - \left(R_0+\frac{E-B_0^2}{c^2}\right)W + \frac{3B_0}{2c^2}W^2 + \frac{1}{2c^2}W^3 - \frac{B_0}{c^2}E = 0.
\end{equation}
Multiplying both sides of Eq.~\eqref{eq: second sub} with $2W_z$ and integrating with respect to $z$ yields,
\begin{equation}\label{eq: first-order ODE}
    \begin{aligned}
    \left(W_z\right)^2 &= -\frac{1}{4c^2}W^4 - \frac{B_0}{c^2}W^3 + \left(R_0+\frac{E-B_0^2}{c^2}\right)W^2 + \frac{2B_0E}{c^2}W + M\\
    &=-\frac{1}{4c^2}(W-W_1)(W-W_2)(W-W_3)(W-W_4),
    \end{aligned}
\end{equation}
where $M$ is another constant of integration, and $W_1 \leq W_2 \leq W_3 \leq W_4$ are the four roots of the potential curve described by the polynomial: 
\begin{equation}\label{eq: Potential curve}
P(W) = W^4 + 4B_0W^3 + \left(-4R_0c^2-4\left(E-B_0^2\right)\right)W^2 - 8B_0EW - 4c^2M. 
\end{equation}
The equations that these roots satisfy will be:
\begin{equation}\label{eq: linear system relation}
    \begin{aligned}
        &W_1+W_2+W_3+W_4 = -4B_0,\\
        &W_1W_2+W_1W_3+W_1W_4+W_2W_3+W_2W_4+W_3W_4 = -4\left(R_0c^2+E-B_0^2\right),\\
        &W_1W_2W_3+W_1W_2W_4+W_1W_3W_4+W_2W_3W_4=8B_0E,\\
        &W_1W_2W_3W_4 = -4c^2M.
    \end{aligned}
\end{equation}

\begin{figure}[t!]
    \centering
    \includegraphics[width=0.32\linewidth]{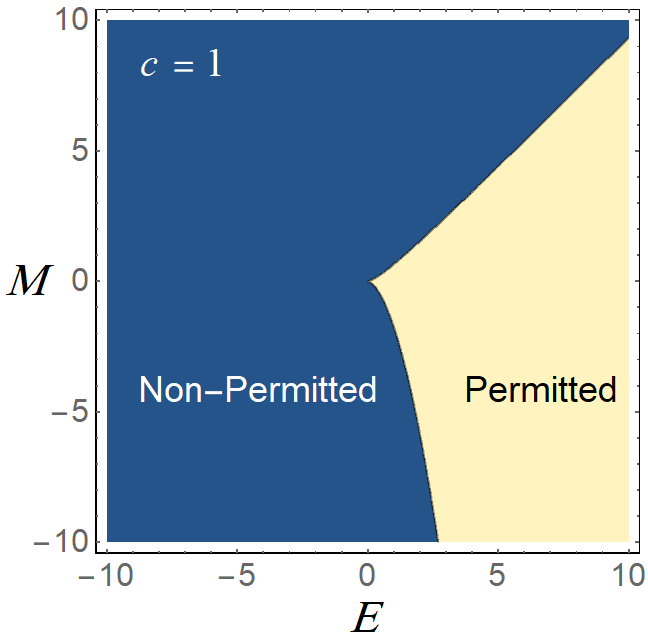}\hspace{0.2cm}
    \includegraphics[width=0.32\linewidth]{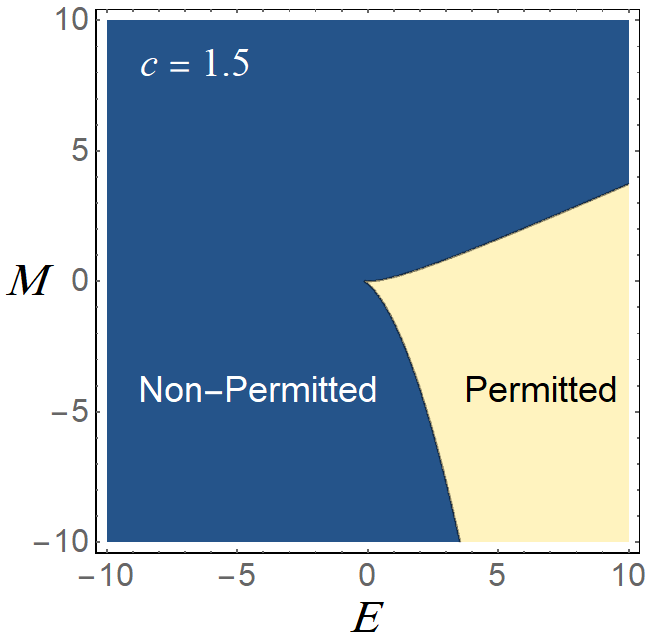}\hspace{0.2cm}
    \includegraphics[width=0.32\linewidth]{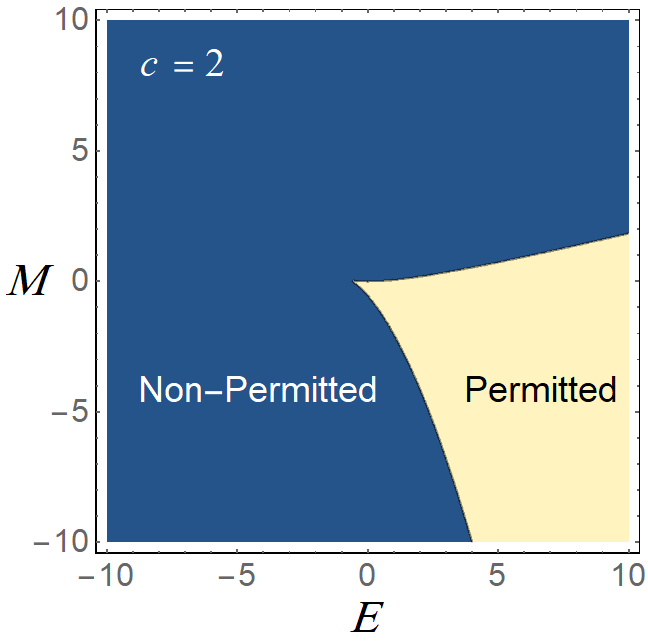}
    \caption{Permitted (light-colored) and non-permitted (dark-colored) regions in the $(E, M)$-parametric space, for three choices of the speed, namely $c=1, 1.5, 2$.}
    \label{fig:roots property analysis}
\end{figure}

The number of real roots $W_{1,2,3, 4}$ plays a crucial role in the existence and the physical relevance of these solutions. In order for the periodic solution 
(that plays a central role in the DSW waveform) to be physically relevant, 
the corresponding inverse carrier density $R$ must remain positive throughout the associated oscillation.
This, however, cannot happen when  \eqref{eq: Potential curve} has no
real solutions, while when there are only two, the smallest one corresponds to negative values of $R$. Thus, the permitted $(E,M)$-parametric region of the problem is restricted to the 
case in which there are four real solutions of \eqref{eq: Potential curve}. In 
Fig.~\ref{fig:roots property analysis} we show the permitted and 
forbidden (i.e., producing physically inadmissible solutions) 
parametric regions for three characteristic values of the speed, namely for $c=1$, $1.5$ and $2$;
in the relevant domain, the positivity constraint on $R$ has also been
taken into account. We note that $E$ must be positive in all cases, while the permitted region shrinks for larger values of $c$. For the
interested reader, a more detailed algebraic analysis
of the polynomial roots is provided in the Appendix.

When the periodic oscillation occurs in $W_3 \leq W \leq W_4$, a direct integration of the ODE in Eq.~\eqref{eq: first-order ODE} yields the following periodic traveling-wave solution to the AA model, expressed in terms of the Jacobi elliptic functions,
\begin{equation}\label{eq: periodic soln}
    W(z) = W_3 + \frac{\left(W_4-W_3\right)\text{cn}^2\left(\zeta, m\right)}{1 + \frac{W_4-W_3}{W_3-W_1}\text{sn}^2\left(\zeta,m\right)},
\end{equation}
where 
\begin{equation}
   \begin{aligned}
    &\zeta = \frac{\sqrt{|\mu|(W_3-W_1)(W_4-W_2)}z}{2},\\
    &m = \frac{(W_4-W_3)(W_2-W_1)}{(W_4-W_2)(W_3-W_1)},
    \end{aligned}
\end{equation}
and $\mu = -\frac{1}{4c^2}$.

\section{Whitham modulation equations}

We can derive the Whitham modulation equations for the AA model represented by the system \eqref{eq: AA model}. Firstly, we notice that the Lagrangian density for Eq.~\eqref{eq: AA model} is known~\cite{PhysRevE.106.034209}:
\begin{equation}\label{eq: Lagrangian density}
    \mathbb{L} = \frac{1}{2}\rho_t^2 + \frac{1}{2}w_x^2 + \frac{1}{2}R_0w^2 + \frac{1}{2}\rho_xw^2 + B_0\rho_xw,
\end{equation}
where $u = \rho_x$. 

Meanwhile, we notice that a substitution of the relation $u = \rho_x$ into the AA system \eqref{eq: AA model} yields the following equations,
\begin{equation}\label{eq: AA model on another level}
    \begin{aligned}
        &\rho_{tt} + B_0w_x + \frac{1}{2}\left(w^2\right)_x = 0,\\
        &w_{xx} - R_0w - B_0\rho_x - \rho_x w = 0.
    \end{aligned}
\end{equation}

To derive the associated Whitham modulation system for the AA model \eqref{eq: AA model}, we apply the so-called Whitham's averaged-Lagrangian 
method~\cite{whitham2011linear}.
To this end, we first notice that, for the periodic traveling waves of the AA model \eqref{eq: AA model}, the behaviors of the two fields of $\rho, w$ 
will generally assume the form,
\begin{equation}\label{eq: behavior of periodic soln}
    \rho = \beta x - \gamma t + \Psi(\theta), \quad
    w = \varphi(\theta); \quad \theta = kx - \omega t,
\end{equation}
where $\beta, \gamma$ are two arbitrary constants, and both $\Psi(\theta)$ and $\varphi(\theta)$ are assumed to be periodic functions with a fixed period $2\pi$. Importantly, it is worth noting that the pair $(\beta,\gamma)$ is the counterpart of $(k,\omega)$ and we define $\eta = \beta x - \gamma t$ as the pseudo-phase.

We then insert Eq.~\eqref{eq: behavior of periodic soln} into system \eqref{eq: AA model on another level} to obtain that, 
\begin{equation}\label{eq: After sub}
    \begin{aligned}
        &\Psi_\theta = \frac{1}{\omega^2}\left(A - B_0k\varphi - \frac{k}{2}\varphi^2\right),\\
        &\rho_t = -\gamma - \frac{A}{\omega} + \frac{B_0}{v_p}\varphi + \frac{1}{2v_p}\varphi^2,\\
        &\rho_x = \beta + \frac{A}{\omega v_p} - \frac{B_0}{v_p^2}\varphi - \frac{1}{2v_p^2}\varphi^2, \\
        &k^2\left(\varphi_\theta\right)^2 = -\frac{1}{4v_p^2}\varphi^4 - \frac{B_0}{v_p^2}\varphi^3 - \left(-R_0+\frac{B_0^2}{v_p^2}-\beta-\frac{A}{\omega v_p}\right)\varphi^2 + 2\left(B_0\beta+\frac{AB_0}{\omega v_p}\right)\varphi + F \equiv G\left(\varphi\right),
    \end{aligned}
\end{equation}
where $A,F$ are two constants of integration, and $v_p = \omega/k$ denotes the phase speed.

In addition, we note that the $2\pi$ periodicity in the phase $\theta$ implies the following relations,
\begin{equation}\label{eq: two important rela}
    \frac{2\pi}{k} = \oint\frac{d\varphi}{\sqrt{G(\varphi)}}, \quad 0 = \oint\Psi_\theta d\theta,
\end{equation}
which are referred to as the nonlinear dispersion relation of the model \eqref{eq: AA model} {and the $2\pi$ period constraint}.
Then, substitution of Eq.~\eqref{eq: After sub} into the Lagrangian density in Eq.~\eqref{eq: Lagrangian density} yields,
\begin{equation}
    \mathbb{L} = k^2\left(\varphi_\theta\right)^2 + \left(A+\omega\gamma\right)\Psi_\theta + \frac{1}{2}\left(\gamma^2-F-\frac{A^2}{\omega^2}\right).
\end{equation}
With the relations derived in Eq.~\eqref{eq: two important rela}, the averaged Lagrangian is then computed as follows,
\begin{equation}\label{eq: Averaged Lagrangian}
    \begin{aligned}
    \mathcal{L}\left(k,\omega,\beta,\gamma,A,F\right) &= \frac{1}{2\pi}\int_0^{2\pi}\mathbb{L}d\theta \\
    &= kI\left(A,F,v_p\right) + \frac{1}{2}\left(\gamma^2-F-\frac{A^2}{\omega^2}\right),
    \end{aligned}
\end{equation}
where $I\left(A,F,v_p\right)$ is defined as the following "action integral",
\begin{equation}\label{eq: Action integral}
    I\left(A,F,v_p\right) = \frac{1}{2\pi}\oint\sqrt{G(\varphi)}d\varphi. 
\end{equation}
Now, we notice that there are  six parameters in total $\Theta = \left(\omega,k,\beta,\gamma,A,F\right)$ in our averaged Lagrangian density \eqref{eq: Averaged Lagrangian}. However, it is important to observe that these six parameters are not all independent, and indeed, we shall see how such a parameter space is reduced in the process of deriving the Whitham modulation equations.

To describe the slow modulation of all the parameters of the periodic traveling waves and derive the associated modulation equations, we first need to introduce the slow spatial and temporal variables:
\begin{equation}\label{eq: Slow variables}
    X = \epsilon x, \quad T = \epsilon t,
\end{equation}
where $0 < \epsilon \ll 1$ is a small number. 

Then, we treat all six parameters in $\Theta$ as functions of the two slow variables $X, T$ (i.e., $\Theta = \Theta(X,T)$). In addition, we define the following fast generalized phase and pseudo-generalised phase: $\theta = \epsilon^{-1}S(X,T)$ and $\eta = \epsilon^{-1}\widetilde{S}(X,T)$, respectively. By the chain rule, we must have the following relations,
\begin{equation}\label{eq: key relations}
    \begin{aligned}
        &k(X,T) = S_X, \quad -\omega(X,T) = S_T,\\
        &\beta(X,T) = \widetilde{S}_X, \quad -\gamma(X,T) = \widetilde{S}_T.
    \end{aligned}
\end{equation}
Then, using  the compatibility conditions of $S_{XT} = S_{TX}$ and $\widetilde{S}_{XT} = \widetilde{S}_{TX}$, we obtain the following two equations,
\begin{equation}\label{eq: conservation of waves}
    k_T + \omega_X = 0, \quad \beta_T + \gamma_X = 0,
\end{equation}
which are the so-called equations of conservation of waves. 

The remaining modulation equations will follow from the following 
averaged-Lagrangian variational principle,
\begin{equation}\label{eq: var pri}
   \delta\int_{-\infty}^{\infty}\int_{-\infty}^{\infty}\mathcal{L}\left(k,\omega,\beta,\gamma,A,F\right)dXdT = 0,
\end{equation}
which yields the Euler-Lagrange equations:
\begin{equation}\label{eq: e-l eqns from var prin}
    \begin{aligned}
        &\delta A: \mathcal{L}_A = 0, \quad \delta F: \mathcal{L}_F = 0,\\
        &\delta S: \frac{\partial}{\partial T}\mathcal{L}_\omega - \frac{\partial}{\partial X}\mathcal{L}_k = 0,\\
        &\delta \widetilde{S}: \frac{\partial}{\partial T}\mathcal{L}_\gamma - \frac{\partial}{\partial X}\mathcal{L}_\beta = 0.
    \end{aligned}
\end{equation}
We note that the first two equations in \eqref{eq: e-l eqns from var prin} yield,
\begin{equation}\label{eq: first two variations}
    \omega^2kI_A = A, \quad 2kI_F = 1.
\end{equation}
These two relations reduce the original six parameters  to four independent ones. As a result, we need four equations to form a closed modulation system. Indeed, between the last two Euler-Lagrange equations in system \eqref{eq: e-l eqns from var prin}, and the two conservation of waves equations in \eqref{eq: conservation of waves}, we obtain four modulation equations which together form a closed modulation system. Because of the complexity of the last two modulation equations from Eq.~\eqref{eq: e-l eqns from var prin}, we will not display them here. While we will not explore the full set of Whitham modulation
equations further, in what follows we will examine the (simpler)
dispersionless limit of the model, which, in turn, will enable
us to analyze the leading and the trailing edge dynamics
of the DSW, through the tools of the well-established~\cite{10.1063/1.1947120}
DSW fitting method.

\section{Riemann problem and jump condition}\label{sec: jump condition}

We can rewrite the evolution equations in Eq.~\eqref{eq: AA model} as the following $p$-system.
\begin{equation}\label{eq: equivalent p system}
    \begin{aligned}
        &u_t = v_x,\\
        &v_t = -B_0w_{x} - \frac{1}{2}\left(w^2\right)_{x},
    \end{aligned}
\end{equation}
which is subject to the constraint:
\begin{equation}\label{eq: constraint eq}
    w_{xx} - R_0w - B_0u - uw = 0.
\end{equation}
To find the dispersionless system associated with Eqs.~\eqref{eq: equivalent p system}--\ref{eq: constraint eq}, we neglect the 
dispersion associated with the term 
involving $w_{xx}$ in the constraint equation \eqref{eq: constraint eq}, so that
\begin{equation}
    u = -\frac{R_0w}{B_0+w}.
\end{equation}
Then, the associated dispersionless system reads
\begin{equation}\label{dispersionless system}
    \begin{aligned}
    &\left[F(w)\right]_t = v_x,\\
    &v_t = -B_0w_x - \frac{1}{2}(w^2)_x,
    \end{aligned}
\end{equation}
where
\begin{equation}\label{def. of F(w)}
    F(w) = -\frac{R_0w}{B_0+w}.
\end{equation}
The dispersionless system in Eq.~\eqref{dispersionless system} can be further cast into the following Riemann invariant form
\begin{equation}\label{e: diagonal system}
    \frac{\partial r_{\pm}}{\partial t} + \lambda_{\pm}\frac{\partial r_{\pm}}{\partial x} = 0,
\end{equation}
with the Riemann invariants of $r_{\pm}$ and the associated characteristic speeds of $\lambda_{\pm}$ defined as follows.
\begin{equation}\label{RI and char speeds}
    \begin{aligned}
        r_{\pm} = v \pm 2\sqrt{R_0B_0}\sqrt{B_0+w}; \quad \lambda_{\pm} = \pm \frac{\left(B_0+w\right)^{3/2}}{\sqrt{R_0B_0}}.
    \end{aligned}
\end{equation}
\begin{figure}[t!]
    \centering
    \includegraphics[width=0.8\linewidth]{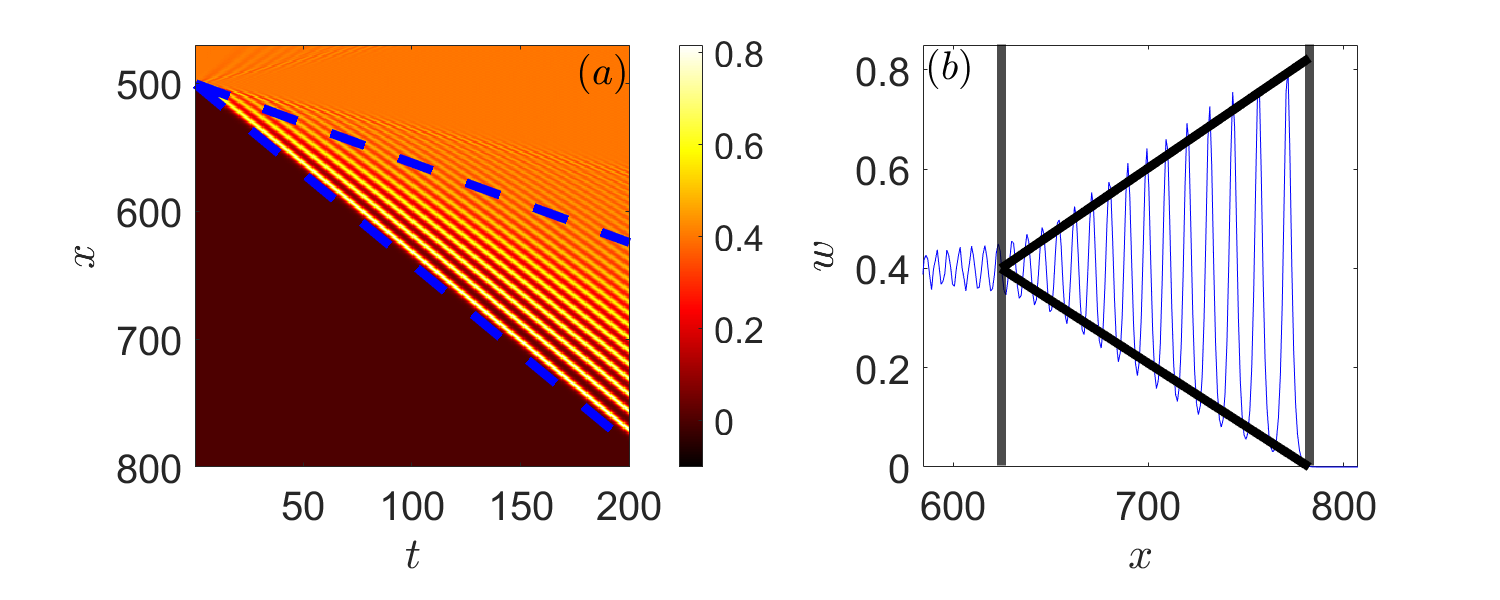}
    \caption{The dispersive shock wave of the AA model \eqref{eq: AA model}. The panel $(a)$ depicts the space-time evolution of the DSW, where the two dashed blue lines represent the theoretical predictions on the linear and solitonic edges of the DSW based on the DSW-fitting results. $(b)$ shows the spatial profile of the DSW at $t = 200$ with  background value $w_- = 0.4$.}
    \label{fig: Simulation of the Riemann problem of AA model}
\end{figure}
To set up an appropriate set of initial conditions for the system in Eq.~\eqref{eq: equivalent p system} {for the observation of right-propagating waves},  we  fix the Riemann invariant of $r_-$ while varying $r_+$ so that 
\begin{equation}\label{e: fixed RI}
    r_-(w_-,v_-) = r_-(w_+,v_+),
\end{equation}
where $w_{\pm}$ and $v_{\pm}$ refer to the background values applied in the following initial data
\begin{equation}\label{eq: ICs for w and v}
    w(x,0) = \begin{cases}
        w_-, \quad a < x < b,\\
        w_+, \quad x \leq a \vee x \geq b,
    \end{cases}
    \quad
    v(x,0) = \begin{cases}
        v_-, \quad a < x < b,\\
        v_+, \quad x \leq a \vee x \geq b.
    \end{cases}
\end{equation}
{It is worthwhile to note that if one instead fixes the Riemann invariant of $r_+$ while varying $r_-$, then the associated constructed initial data are expected to lead to the formation of left-propagating waves.}

We note that by substituting $r_-$ specified in Eq.~\eqref{RI and char speeds} 
into Eq.~\eqref{e: fixed RI}, we arrive at
\begin{equation}\label{eq: jump condition}
    v_- - 2\sqrt{R_0B_0}\sqrt{B_0+w_-} = v_+ - 2\sqrt{R_0B_0}\sqrt{B_0+w_+}.
\end{equation}
Eq.~\eqref{eq: jump condition} specifies the necessary jump condition \cite{EL201611} that the initial conditions must satisfy.

Now, we notice that the jump condition \eqref{eq: jump condition} demonstrates the relation between the four background values of $w_{\pm}$ and $v_{\pm}$, so it indicates that when we fix three of them, say $w_{\pm}$ and $v_+$, the remaining background is then  determined self-consistently
from Eq.~\eqref{eq: jump condition}. With this knowledge, we are ready to construct a set of consistent initial conditions to be used for the numerical simulation of Eq.~\eqref{eq: equivalent p system}. 

We propose the following ``box-type" initial data for the variable  $w$.
\begin{equation}\label{eq: box-type for w}
    w(x,0) = w_+ - \frac{1}{2}\left(w_+-w_-\right)\left[\tanh\left(\delta(x-x_l)\right) - \tanh\left(\delta(x-x_r)\right)\right],
\end{equation}
where $x_l$ and $x_r$ refer to the left and right locations of the ``box", respectively, and the parameter $\delta > 0$ characterizes the smoothness of the two jumps occurring at the left and right locations of the box;
i.e., the quantity $1/\delta$ gives a measure of the width of the region
over which the jump in the Riemann problem initial data takes place
(for practical computational purposes). Accordingly, we should expect a sharp jump when $\delta$ is large. Clearly, we can see that the initial data in Eq.~\eqref{eq: box-type for w} is a smoothed version (for the numerical
implementation) of the one specified in Eq.~\eqref{eq: ICs for w and v}.

On the other hand, for the variable  $v$, we choose the following proposed initial data so that it is consistent  with the jump condition in Eq.~\eqref{eq: jump condition}.
\begin{equation}\label{eq: IC for v}
    v(x,0) = 2\sqrt{R_0B_0}\left(\sqrt{B_0+w(x,0)} - \sqrt{B_0+w_+}\right) + v_+.
\end{equation}
To see this, we note that the initial condition in \eqref{eq: IC for v} implies
\begin{equation}
    v_- = 2\sqrt{R_0B_0}\left(\sqrt{B_0+w_-} - \sqrt{B_0+w_+}\right) + v_+,
\end{equation}
which is clearly consistent with Eq.~\eqref{eq: jump condition}.

However, we shall ignore the constant $v_+$ in Eq.~\eqref{eq: IC for v} as it does not affect the evolution of the model \eqref{eq: equivalent p system} at all. In addition, since the system \eqref{eq: equivalent p system} evolves the variable $u$ in time, we need initial data for $u$ as well, which shall be, according to the constraint condition specified in Eq.~\eqref{eq: constraint eq},
\begin{equation}
    \begin{aligned}
    u(x,0) &= \frac{\left(w_0(x)\right)_{xx} - R_0w_0(x)}{B_0 + w_0(x)},\\
    &= \frac{\mathcal{F}^{-1}\left[-k^2\mathcal{F}\left[w_0(x)\right]\right] - R_0w_0(x)}{B_0 + w_0(x)}
    \end{aligned}
\end{equation}
where $w_0(x) = w(x,0)$, while $\mathcal{F}, \mathcal{F}^{-1}$ refer to the Fourier and inverse Fourier transforms, and $k$ here denotes the Fourier wavenumber.

Then, we briefly discuss  the numerical methods used in the simulation of the model in Eq.~\eqref{eq: equivalent p system}. In particular, we apply the RK4 integration scheme for time evolution, and a pseudospectral discretization in space to numerically solve Eq.~\eqref{eq: equivalent p system}. Meanwhile, for the constraint
equation in \eqref{eq: constraint eq}, when we have the solutions for the variable $u$ at time $t$, we solve the following linear system to obtain $w$.
\begin{equation}\label{eq: linear system to solve for w}
    \left[D_2 - \text{diag}\left(R_0+u\right)\right]w = B_0u,
\end{equation}
where $D_2$ denotes the Laplacian matrix, 
and $\text{diag}\left(f\right)$ represents the matrix whose diagonal entries are the entries of the column vector $f$. Fig.~\ref{fig: Simulation of the Riemann problem of AA model} shows the evolution dynamics of the AA model \eqref{eq: AA model} with the initial conditions given in Eqs.~\eqref{eq: box-type for w} and \eqref{eq: IC for v}. It demonstrates the right-propagating DSW since we have fixed the Riemann invariant of $r_-$. Similarly, a left-propagating DSW can be observed if one chooses to instead fix the Riemann invariant of $r_+$. Moreover, we notice that the two dashed blue lines in panel $(a)$ represent the theoretical predictions for the linear and solitonic edges of the AA DSW based on the results in Section \ref{sec: DSW fitting} that will be presented below. The black triangular region in panel $(b)$ of Fig.~\ref{fig: Simulation of the Riemann problem of AA model} also represents the theoretical prediction for the AA DSW spatial profile based on the DSW-fitting results in Section \ref{sec: DSW fitting}. Specifically, we construct such a triangular region as follows. For the upper oblique straight line, we connect the two points of $(n_-, w_-)$ and $(n_+, w_+ + a_+)$, where $n_{\pm}$ denote the DSW-fitting estimated linear and solitonic edges of the AA DSW, and $a_+$ is the DSW-fitting predicted DSW amplitude. On the other hand, the lower line is obtained by connecting the two points of $(n_-,w_-)$ and $(n_+,w_+)$. Finally, the two vertical lines in panel $(b)$ are simply $x = n_-$ and $x = n_+$, respectively.

\section{DSW fitting}\label{sec: DSW fitting}

In this section, we apply the so-called DSW-fitting method to obtain insightful theoretical predictions for the various edge features of the DSW in the AA model \eqref{eq: AA model}. To begin with, notice that it is expected that the Whitham modulation equations in \eqref{eq: e-l eqns from var prin} shall coincide with the dispersionless system of \eqref{eq: AA model} in the averaged p-system form specified in Eq.~\eqref{dispersionless system}, at both linear and solitonic limits \cite{10.1063/1.1947120,EL201611}. 
Applying the DSW fitting technique at the linear edge
of the structure~\cite{10.1063/1.1947120}, we obtain
\begin{equation}\label{linear-edge simple-wave ode}
    \frac{dk}{d\overline{w}} = \frac{\partial_{\overline{w}}\omega_0}{V(\overline{w}) - \partial_k\omega_0}, \quad k(0) = 0,
\end{equation}
where $\omega_0$ denotes the linear dispersion relation of the AA model. The dispersion relation  $\omega_0$ can be obtained by seeking a plane-wave solution in the form of an infinitesimal perturbation to the homogeneous background $\overline{w}$: $w(x,t) = \overline{w} + \eta \exp\left[i(kx-\omega t)\right]$, where $0 < \eta \ll 1$. A direct substitution of this plane-wave ansatz into the AA model \eqref{eq: AA model} yields 
\begin{equation}\label{linear dispersion relation of AA model}
    \omega^2 = \frac{\left(B_0+\overline{w}\right)^3k^2}{B_0\left(R_0+k^2\right)+\overline{w}k^2}.
\end{equation}
We then take the positive branch (with positive speed
for the right-moving DSWs) so that
\begin{equation}\label{positive-branch dispersion relation}
    \omega_0 = \frac{(B_0+\overline{w})^{3/2}k}{\sqrt{B_0(R_0+k^2) + \overline{w}k^2}}.
\end{equation}
Then, we note that the initial condition of $k(0) = 0$ in Eq.~\eqref{linear-edge simple-wave ode} is imposed because the wavenumber $k$ at the solitonic edge of the DSW is zero. Then, $V(\overline{w}) = \lambda_+(\overline{w})$ since we are interested only in the right-propagating DSW and the second Riemann invariant of $r_-$ is fixed to be constant, as specified in  section \ref{sec: jump condition}. Then, we need to integrate the initial-value problem \eqref{linear-edge simple-wave ode} up to $\overline{w} = w_-$ to obtain the linear-edge wavenumber defined as $k_- = k(w_-)$. Moreover, the linear-edge speed, denoted by $s_-$, of the DSW is obtained by computing the group velocity:
\begin{equation}
    s_- = \partial_k\omega_0\left(w_-,k_-\right).
\end{equation}
On the other hand, for the solitonic edge, we introduce the following conjugate dispersion relation
\begin{equation}
    \widetilde{\omega}_s = -i\omega_0\left(\overline{w},i\widetilde{k}\right),
\end{equation}
where $\widetilde{k}$ denotes the conjugate wavenumber \cite{EL201611}.

Then, the simple-wave ODE at the solitonic edge is given as follows.
\begin{equation}\label{eq: solitonic-edge ODE}
    \frac{d\widetilde{k}}{d\overline{w}} = \frac{\partial_{\overline{w}}\widetilde{\omega}_s}{V(\overline{w}) - \partial_{\widetilde{k}}\widetilde{\omega}_s}, \quad \widetilde{k}(w_-) = 0.
\end{equation}
The initial condition $\widetilde{k}(w_-) = 0$ is used for the simple-wave ODE 
because the conjugate wavenumber is zero at the linear edge.

The initial-value problem in Eq.~\eqref{eq: solitonic-edge ODE} needs to be integrated up to $\overline{w} = w_+$ to obtain the value of the conjugate wavenumber at the solitonic edge: $\widetilde{k}_+ = \widetilde{k}(w_+)$. Then, the solitonic-edge speed, denoted as $s_+$, of the DSW can be obtained by computing the associated phase speed:
\begin{equation}\label{eq: solitonic-edge speed}
    s_+ = \frac{\widetilde{\omega}_s}{\widetilde{k}}\left(w_+, \widetilde{k}_+\right).
\end{equation}
Furthermore, based on the DSW-fitting theoretical prediction for the solitonic-edge speed of $s_+$ and the soliton amplitude-speed relation specified in Eq.~\eqref{e: soliton amplitude}, we can also obtain the theoretical prediction for the DSW solitonic amplitude, denoted as $a_+$,
\begin{equation}\label{e: DSW-fitting soliton amplitude}
    a_+ = \frac{2B_0(s_+-\gamma)}{\gamma}.
\end{equation}
This reflects the feature that the leading edge manifests a solitary
traveling wave with the corresponding speed $s_+$.
We will return to the corresponding comparison of the DSW-fitting
predictions in the numerical section below, but we remind the 
reader that an indication of the accuracy of the relevant prediction
is incorporated in the dynamics of Fig.~\ref{fig: Simulation of the Riemann problem of AA model}.

\section{Rarefaction wave}

\begin{figure}[t!]
    \centering
    \includegraphics[width=0.8\linewidth]{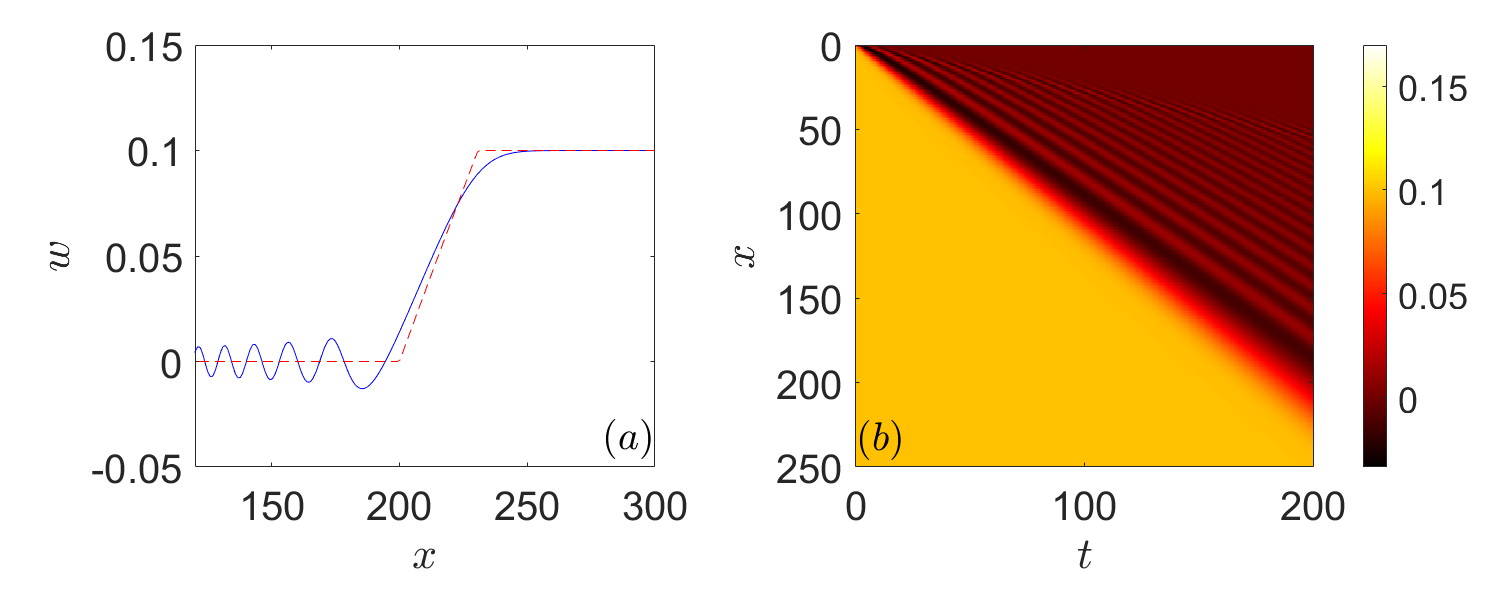}
    \caption{The dynamics of the numerical rarefaction waves of the AA model \eqref{eq: AA model}. The panel $(a)$ depicts the comparison of the self-similar solution \eqref{eq: self-similar soln} (in dashed red) with the numerically observed rarefaction wave (in solid blue) at $t = 200$. Note that $w_- = 0$ and $w_+ = 0.1$. The panel $(b)$ displays the spatio-temporal evolution dynamics of the numerical rarefaction wave of the AA model in Eq.~\eqref{eq: AA model}.}
    \label{fig: RW comparison}
\end{figure}

We can also theoretically study the numerical rarefaction wave observed in the evolution of the Riemann problem with the initial data specified in Eq.~\eqref{eq: ICs for w and v}. In particular, the right-propagating RW can be determined by looking for a self-similar solution in the following form:
\begin{equation}\label{eq: self-similar ansatz}
    r_+(x,t) = r_+(\kappa), \quad \kappa = \frac{x}{t}.
\end{equation}
Substitution of this ansatz \eqref{eq: self-similar ansatz} into the system \eqref{e: diagonal system} yields the following condition for the self-similar solution:
\begin{equation}
    \lambda_+(w,v) = \kappa.
\end{equation}
Then, the self-similar solution in terms of $w$ reads
\begin{equation}\label{eq: self-similar soln}
    w(x,t) = \begin{cases}
        w_-, \quad x \leq t\lambda_+\left(w_-,v_-\right),\\
        (x/t)^{2/3}\left(R_0B_0\right)^{1/3} - B_0, \quad t\lambda_+(w_-,v_-) < x < t\lambda_+(w_+,v_+),\\
        w_+, \quad x \geq t\lambda_+(w_+,v_+).
    \end{cases}
\end{equation}
Panel $(a)$ in Fig.~\ref{fig: RW comparison} shows the comparison of the self-similar solution in Eq.~\eqref{eq: self-similar soln} with the numerical rarefaction wave at $t = 200$. The close alignment of the two solutions indicates a good approximation of the self-similar solution \eqref{eq: self-similar soln}. {In addition, it is worth noting  the linear oscillations occurring within the window of $x\in \left[120,200\right]$, as depicted in panel $(a)$ of Fig.~\ref{fig: RW comparison}. In particular, these small-amplitude linear oscillations are attributed to radiation emitted by the tail of the rarefaction wave.
}

\section{Numerical validation}\label{sec: numerical validation}

In this section, we perform numerical comparisons between the theoretical DSW-fitting predictions for distinct edge characteristics of the AA DSW and their associated numerical counterparts. However, before we discuss these relevant comparisons, we introduce briefly some necessary preliminaries including our methods applied to numerically compute the edge speeds of the DSW.

Firstly, to compute the numerical solitonic-edge speed of the AA DSW, we keep track of the $x$ coordinate of the highest peak of the DSW, denoted as $x_+$.
Accordingly, we obtain a time-series of $x_+(t)$, and we treat $x_+(t)$ as the location of the solitonic edge of the DSW. Then, the numerical DSW solitonic-edge speed, denoted by $s_+$, can be readily approximated as follows:
\begin{equation}\label{numerical solitonic-edge speeds}
    s_+ = \frac{x_+(t_2) - x_+(t_1)}{t_2 - t_1},
\end{equation}
where $t_1, t_2$ are two (proximal) time snapshots during the simulation. However, it is important to notice that the choice of $t_1$ and $t_2$ should be influenced by the development of the DSW. In particular, only once the DSW is fully developed (i.e., the amplitude of the DSW ceases to increase), say at time $t_f$, can we pick two nearby time snapshots so that $t_1, t_2 \ge t_f$. 

Next, for the linear-edge speed denoted by $s_-$, we can simply replace $x_+$ in Eq.~\eqref{numerical solitonic-edge speeds} by $x_-$ which refers to the linear-edge location of the AA DSW to obtain the speed:
\begin{equation}\label{linear-edge speed}
    s_- = \frac{x_-(t_2) - x_-(t_1)}{t_2 - t_1}.
\end{equation}
However, it is much more challenging to measure the linear-edge location of $x_-(t)$. Although different approaches can be used to numerically compute $x_-(t)$, here we reconstruct it as follows. We define the following two quantities:
\begin{equation}\label{label quantities}
    \Upsilon^u = w_- + \frac{\left|w_--w_+\right|}{M}, \quad
    \Upsilon^l = w_- - \frac{\left|w_--w_+\right|}{M},
\end{equation}
where $M \in \mathbb{N}$ denotes an integer (e.g., $M = 5$). 
Then, based on the two quantities in Eq.~\eqref{label quantities}, we can construct the following two intervals:
\begin{equation}\label{eq: two intervals}
    I^u = \left(\Upsilon^u - \gamma, \Upsilon^u + \gamma\right), \quad I^l = \left(\Upsilon^l - \gamma, \Upsilon^l + \gamma\right),
\end{equation}
where $\gamma > 0$ is a small number, typically set to $\gamma = \frac{\left|w_--w_+\right|}{5}$. Then we collect all the local maxima and minima within the two intervals of $I^u$ and $I^l$. Using these local extrema, we fit them with two straight lines using a least-square based method so that the two lines intersect at a point, and we then regard the intersection as the linear-edge location $x_-$ of the DSW. We repeat this process for multiple time snapshots so that we can finally obtain a time-series data $x_-(t)$ of the linear-edge location. 

\begin{figure}[t!]
    \centering
    \includegraphics[width=0.8\linewidth]{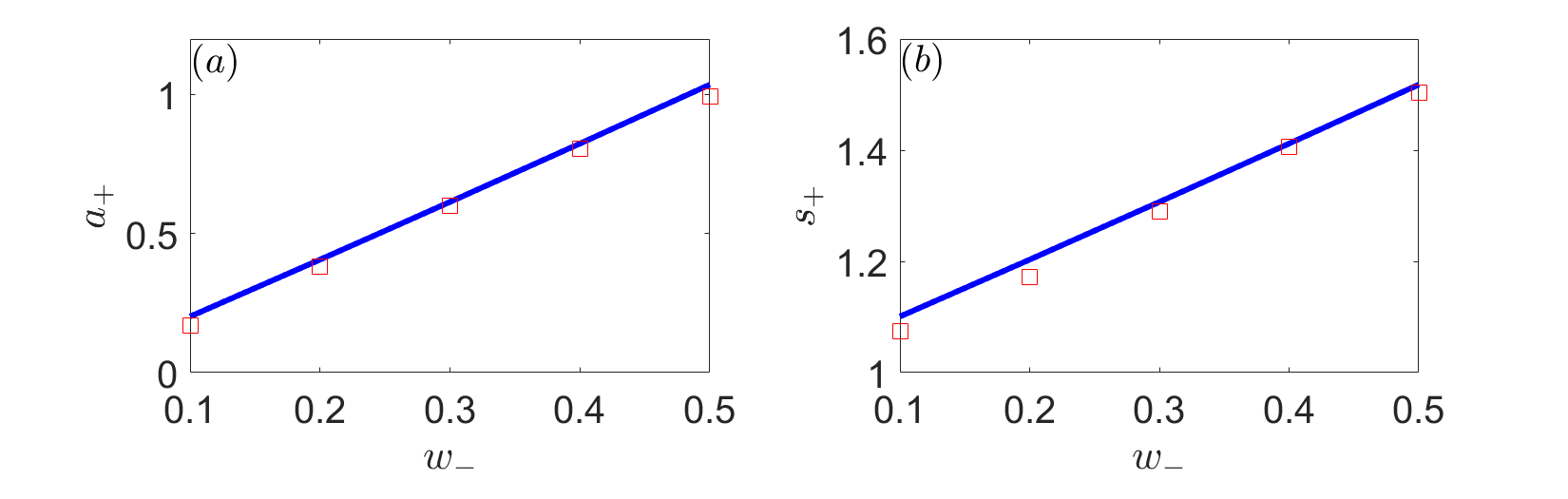}
    \caption{The comparison of the solitonic-edge DSW features. The panel $(a)$ depicts the comparison of the solitonic-edge amplitudes of the DSW while $(b)$ shows that of the solitonic-edge speeds. The blue curves and discrete red squares represent the DSW-fitting theoretical predictions and the numerically estimated counterparts, respectively. Note also that $w_+$ is always fixed to be zero, while the values of $w_-$ are varied within the interval of $\left[0.1,0.5\right]$.}
    \label{fig:solitonic-edge comparison}
\end{figure}

\begin{figure}[b!]
    \centering
    \includegraphics[width=0.4\linewidth]{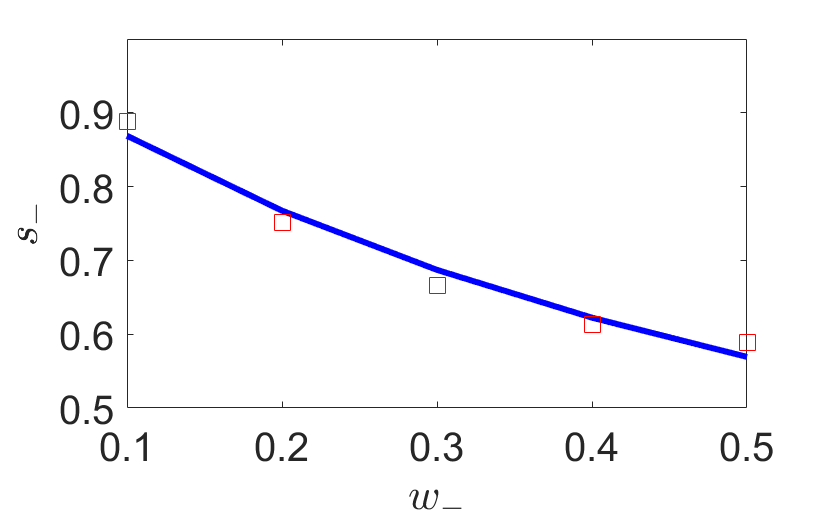}
    \caption{The comparison of the linear-edge DSW speeds. The blue curve and red squares refer to the DSW-fitting theoretical prediction and the numerically computed counterparts, respectively.}
    \label{fig:linear-edge comparison}
\end{figure}

Now, we are ready to discuss the relevant numerical comparisons. 
As discussed before, a typical example of a DSW comparison with the
DSW-fitting outcomes for a particular parameter set is given in 
Fig.~\ref{fig: Simulation of the Riemann problem of AA model}.
Upon varying the parameter $w_-$,
 Figs.~\ref{fig:solitonic-edge comparison}-\ref{fig:linear-edge comparison} depict the comparisons of the DSW-fitting theoretical predictions on the edge speeds (in solid blue curves) of the DSW and the numerically measured edge speeds (in discrete red squares). We can clearly see that for both the solitonic and linear-edge speeds, the discrete red squares are located very close to the blue curves, indicating that the DSW-fitting method provides 
accurate predictions for the edge speeds 
 of the AA DSW, as well as for the amplitude of its leading edge.
 This suggests that the methodology deployed herein is suitable 
 for providing a systematic characterization of the AA DSW features.
 Notice that the prediction remains accurate for a considerable range
 of jumps associated with Riemann initial data (and not only for small
 jumps).

\section{KdV reduction}

As a final element of our analysis, 
in this section, we demonstrate another approach to analyze AA DSWs through a  multiscale approach, namely the KdV reduction of the AA model which, as discussed above, is a valuable tool for approximating
features of the AA model (previously used, e.g., to approximate the solitary 
waves of AA via those of KdV in~\cite{NAIRN_BINGHAM_ALLEN_2005,PhysRevE.102.013209}). To this end, we introduce the following slow spatial and temporal variables \cite{PhysRevE.102.013209}
\begin{equation}\label{KdV reduction to AA model: change of variables}
    X = \epsilon^{1/2}\left(x - Ct\right), \quad T = \epsilon^{3/2}t,
\end{equation}
and the asymptotic expansions for the variables $u$ and $w$:
\begin{equation}\label{asymptotic expansions}
   u = \epsilon u_1 + \epsilon^2 u_2 + \ldots, \quad w = \epsilon w_1 + \epsilon^2 w_2 + \ldots
\end{equation}
where $C^2 \equiv \frac{B_0^2}{R_0}$.

Substituting Eqs.~\eqref{KdV reduction to AA model: change of variables}-\eqref{asymptotic expansions} into the AA model in Eq.~\eqref{eq: AA model} yields, at leading order, $\mathcal{O}(\epsilon)$:
\begin{equation}\label{Leading-order relation}
    u_1 = -\frac{R_0}{B_0}w_1.
\end{equation}
The KdV equation is obtained by collecting relevant terms at 
the  order of $\mathcal{O}(\epsilon^3)$
which reads
\begin{equation}\label{KdV reduction of AA}
    w_{1T} + \frac{3B_0}{2CR_0}w_1w_{1X} + \frac{C}{2R_0}w_{1XXX} = 0.
\end{equation}
We shall now approximate the AA DSW leveraging that of the KdV reduction \eqref{KdV reduction of AA}. To this end, we first note that the KdV equation in \eqref{KdV reduction of AA} admits the following traveling solitary-wave solution
\begin{equation}\label{KdV-reduction soliton}
    w_1(X,T) = \frac{2CR_0v}{B_0}\text{sech}^2\left(\frac{1}{2}\sqrt{\frac{2R_0v}{C}}\left(X-vT-x_0\right)\right),
\end{equation}
where $v$ is the propagation speed of the KdV solitary wave, and $x_0$ denotes the arbitrary initial phase of the soliton. 

Hence, the soliton amplitude-speed relation is given as
\begin{equation}\label{amplitude-speed relation}
    a = \frac{2CR_0v}{B_0},
\end{equation}
where $a$ denotes the amplitude of the solitary wave
and $v$ its speed. We can then predict both the amplitude and solitonic-edge speed of the DSW based on the DSW-fitting of the KdV reduction \eqref{KdV reduction of AA} as described in section \ref{sec: DSW fitting}. On the one hand, the DSW-fitting prediction for the solitonic-edge speed of the DSW of the KdV reduction in Eq.~\eqref{KdV reduction of AA} reads
\begin{equation}\label{DSW-fitting for s_+}
    s_+^{(X,T)} = \frac{B_0w_{1-}}{R_0C},
\end{equation}
where $w_{1-}$ denotes the left background of the box-type initial condition specified in Eq.~\eqref{unit-jump IC for KdV reduction}.

We notice that the superscript $(X,T)$ is used to emphasize the fact that 
this theoretical prediction for the edge speed is made in the coordinates $(X,T)$. However, since our purpose is to utilize the KdV reduction to gain information on the AA DSW, we ought to transform the solitonic-edge speed $s_+^{(X,T)}$ \eqref{DSW-fitting for s_+} of the KdV DSW from the coordinates of $(X,T)$ back into $(x,t)$. Such transformation leads to
\begin{equation}\label{speed in (x,t)}
    s_+^{(x,t)} = \epsilon s_+^{(X,T)} + C,
\end{equation}
where $s_+^{(x,t)}$ is defined as the solitonic-edge speed associated with the $(x,t)$ coordinates.

\begin{figure}[t!]
    \centering
    \includegraphics[width=0.8\linewidth]{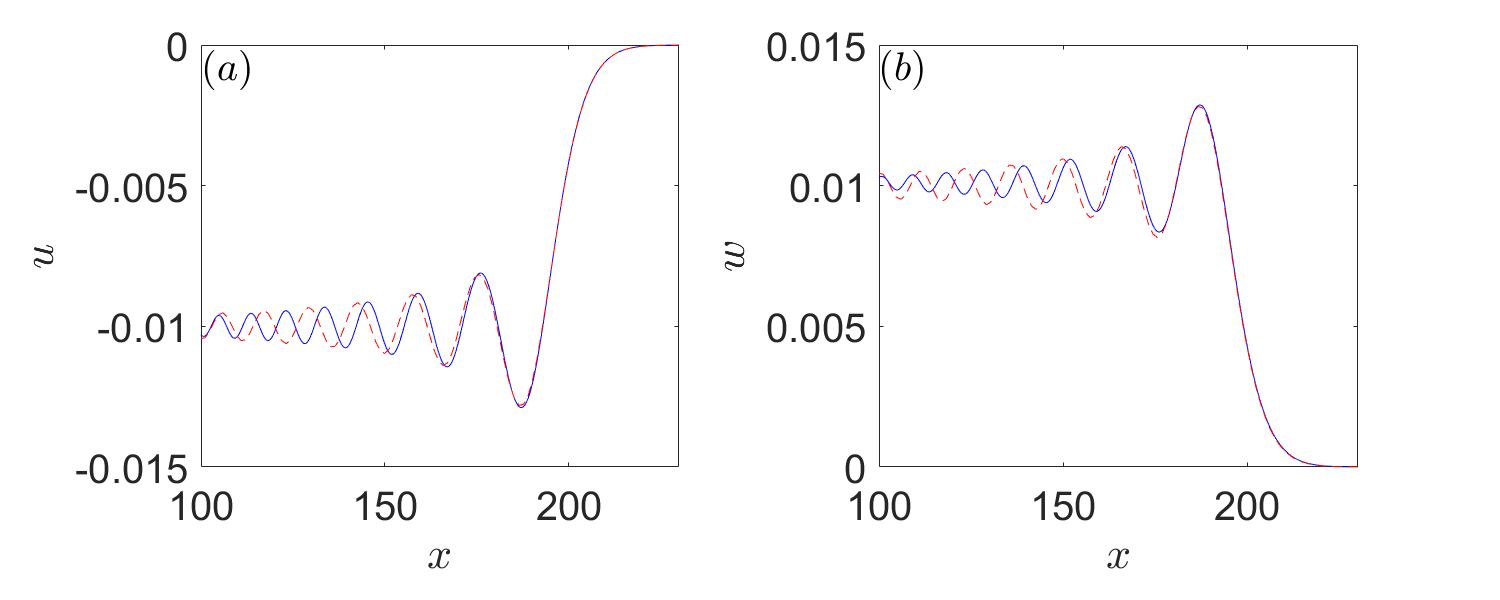}
    \caption{The comparison of the AA (in blue curves) and KdV (in red dashed curves) DSWs  at $t = 200$. The panel $(a)$ depicts the comparison of the DSWs in terms of the $u$ variable, while the panel $(b)$ shows the one in terms of $w$.}
    \label{fig:AA and KdV DSWs comparison}
\end{figure}

Meanwhile, based on the amplitude-speed equality in Eq.~\eqref{amplitude-speed relation} and the relation that $w = \epsilon w_1$, we can also compute the DSW-fitting prediction on the amplitude of the DSW which reads $a_+^{(x,t)} = 2w_-$, where $w_-$ is the background state specified in the initial condition \eqref{eq: box-type for w}.

Finally, we can compare the KdV and AA DSWs through numerical simulations. We recall that since the AA model \eqref{eq: AA model} is a second-order (in time) model, we need initial conditions not only for $u$ and $w$, but also for the ``velocity" variable, which is defined as $s(x,t) \equiv u_t(x,t)$, as well. To construct a consistent set of initial conditions utilized for both the AA and KdV models, we start from the KdV reduction by considering the ``box-type" initial condition for the field variable $w_1$:
\begin{equation}\label{unit-jump IC for KdV reduction}
    w_1(X,0) = w_{1+} - \frac{1}{2}\left(w_{1+} - w_{1-}\right)\left[\tanh\left(\delta(X-X_l\right) - \tanh\left(\delta(X-X_r)\right)\right],
\end{equation}
Then, the initial conditions for the AA variables  $u$ and $w$  can be constructed according to
\begin{equation}
    \begin{aligned}
    &w(x,0) = \epsilon w_1(X,0), \\
    &u(x,0) = \epsilon u_1(X,0) = -\frac{\epsilon R_0}{B_0}w_1(X,0), 
    \end{aligned}
\end{equation}
where now $X = \epsilon^{1/2}x$ at time $t=0$.

Regarding the initial condition for the velocity $s$, we first observe that
\begin{equation}
    \begin{aligned}
        s(x,t) &= \partial_tu(x,t), \\
        &=\epsilon\partial_tu_1(X,T), \\
        &= \epsilon\left(u_{1X}\frac{\partial X}{\partial t} + u_{1T}\frac{\partial T}{\partial t}\right) \\
        &= -C\epsilon^{3/2}u_{1X} + \epsilon^{5/2}u_{1T} \\
        &= C\epsilon^{3/2}\frac{R_0}{B_0}w_{1X} + \epsilon^{5/2}\frac{R_0}{B_0}\left(\frac{3B_0}{2CR_0}w_1w_{1X} + \frac{C}{2R_0}w_{1XXX}\right).
    \end{aligned}
\end{equation}
Therefore, the initial condition for $s$ should read
\begin{equation}
    s(x,0) = C\epsilon^{3/2}\frac{R_0}{B_0}w_{1X}(X,0) + \epsilon^{5/2}\frac{R_0}{B_0}\left(\frac{3B_0}{2CR_0}w_1(X,0)w_{1X}(X,0) + \frac{C}{2R_0}w_{1XXX}(X,0)\right).
\end{equation}
Finally, before we display the results of the numerical comparison, we briefly mention that we applied the ETDRK4 scheme \cite{doi:10.1137/S1064827502410633} for time stepping and a pseudospectral discretization of space to numerically solve the KdV reduction of Eq.~\eqref{KdV reduction of AA}. Fig.~\ref{fig:AA and KdV DSWs comparison} shows the comparison of the AA and KdV DSWs in terms of both the $u$ and $w$ variables. The close alignment of the KdV DSW (shown by the red dashed 
curves) with that of the AA model \eqref{eq: AA model} (blue curves) demonstrates the good performance of the KdV reduction \eqref{KdV reduction of AA} in approximating the AA DSW. In addition, we have computed the numerical solitonic-edge speed of the AA DSW, which is about $0.99$, and also the AA DSW amplitude, which is about $0.01$, based on the method discussed in Sec.~\ref{sec: numerical validation}. These numerically measured edge features agree very well with the corresponding theoretical predictions based on the KdV DSW-fitting results which are $1.0$ and $0.01$ for the solitonic-edge speed and DSW amplitude, respectively. {Moreover, we notice that it is also possible to compare the theoretical KdV DSW with that of the AA model. In particular, the KdV DSW can be analytically constructed~\cite{EL201611} through the (leading order) KdV-Whitham system and the completely integrable structure of the KdV equation. However, we shall avoid utilizing the theoretical KdV DSW to approximate that of the AA model in order to mitigate any additional discrepancy that the approximate nature of the leading order Whitham 
modulation-theory analysis may induce in comparison with the full KdV dynamics.}

\section{Conclusions and future directions}

In the present work we have revisited the Adlam-Allen model
for the description of hydromagnetic waves of cold collisionless 
plasmas. More specifically, we have solved numerically and 
analyzed theoretically the Riemann problem with jump initial
conditions in the context of this model. Our analysis 
has developed the dispersionless limit of the model and explored
the specific predictions of the DSW-fitting methodology for
both the leading and the trailing edges of the DSW. Between those
and the slowly modulated envelope of the periodic traveling
waves (also analyzed), one can obtain a systematic characterization
of the DSW structures. In addition to providing such explicit
predictions for the DSW edge amplitude and speeds, we have also
leveraged the connection with the KdV model to provide an
alternative analytical perspective in the limit of
small-amplitude jumps, where the KdV asymptotic reduction 
remains reasonably valid. We have shown how to carry this out
and how to formulate the comparison with the results  stemming from the KdV 
reduction.
In all the above cases, we have provided comparisons (for
DSWs and rarefaction waves) with the full numerical computations
of the AA model, obtaining good agreement, even when parametrically
varying the height of the Riemann problem jump.

The present work paves the way for further explorations
of dispersive shock waves in plasma settings. For instance, it is
well-known that the AA model is obtained from a reduction of 
Maxwell's equations for a setting in which electrons and ions
in a plasma are subject to a  magnetic field  in one direction,
while no variations in the pertinent fields along the transverse 
directions are considered~\cite{Adlam01051958,JHAdlam_1960}.
It is important to note that numerous variants of such
settings also exist~\cite{Woods_1972}. Exploring the DSW analysis
in the full Maxwell-equation setting would be a particularly
intriguing extension. Additionally, all the considerations
presented herein have been focused on one-dimensional propagation.
Far fewer studies~\cite{EL201611} currently exist for higher-dimensional
settings, which are only beginning to be explored not only theoretically,
but also numerically~\cite{bivolcic}. Extending such multi-dimensional
DSW considerations in plasma settings would be of particular
interest in its own right. Such studies are currently in 
progress and will be reported in future publications.

\paragraph{Acknowledgments:} This work was supported in part by the U.S. National Science Foundation under award PHY-2408988
(P.G.K.). This research was partly conducted while P.G.K. was  visiting the Okinawa Institute of Science and
Technology (OIST) through the Theoretical Sciences Visiting Program (TSVP), the University of
Sydney through the visitor program of the Sydney Mathematical Research Institute (SMRI) and the Department of Mechanical Engineering at Seoul National
University through a Fulbright Fellowship. Their support is gratefully acknowledged.
Finally, this work was also  supported by a grant from the Simons Foundation [SFI-MPS-SFM-00011048, P.G.K.].

\appendix
\section*{Appendix: Polynomial Roots}

As discussed in the main text, for the polynomial $P(W)$, 
we shall restrict the constants of integration 
to be such that $W_{1,2,3,4} \in R$, i.e., that the relevant
polynomial has real roots. This is a necessary, but not sufficient
condition for the physical relevance of the solutions.
Recall (also from the considerations of~\cite{PhysRevE.102.013209}) that
the field $R(x,t)= R_0 + u(x,t)$ represents an inverse density, hence
we need to enforce that $R(x,t)>0$. In what follows, we will assume
that the roots $W_{1,2,3,4}$ satisfy both of these constraints. 
The associated Sturm analysis implies the following facts regarding the polynomial in Eq.~\eqref{eq: Potential curve}: consider the quantities
\begin{equation}
\begin{aligned}
&\Delta = 16A^4C - 4A^3B^2 - 128A^2C^2 + 144AB^2C - 27B^4 + 256C^3,\\
&D = 2A^3 - 8AC + 9B^2,
\end{aligned}
\end{equation}
where
\begin{equation}
    \begin{aligned}
        &A = -2B_0^2 - 4E - 4R_0c^2,\\
        &B = 8B_0R_0c^2,\\
        &C = B_0^4 + 4B_0^2E - 4B_0^2R_0c^2 - 4c^2M.
    \end{aligned}
\end{equation}
Then, the number of real roots of the polynomial of $P(W)$ in Eq.~\eqref{eq: Potential curve} is determined as follows:
\begin{itemize}
    \item If $\Delta < 0$, then $P(W)$ has exactly two real roots (and one complex conjugate pair).
    \item If $\Delta >0$, $A < 0$, and $D < 0$, then $P(W)$ has four distinct real roots.
    \item If $\Delta > 0$, and either $A \geq 0$ or $D \geq 0$, then $P(W)$ has no real roots.
\end{itemize}
Finally, if $\Delta = 0$, then $P(W)$ has at least one multiple root, and a refined Sturm analysis needs to be performed in order to determine the number of distinct real roots.
Now, we notice that in addition to the requirement of real roots of the polynomial in Eq.~\eqref{eq: Potential curve}, we also need to ensure that the field variable $R$ remains physically admissible. This means that we also have to impose the constraint that $U + R_0 > 0$,
which has been taken into account in determining the final admissible set of parameters of $(E,M)$ shown in Fig.~\ref{fig:roots property analysis}.

\bibliography{main}

\bibliographystyle{abbrv}

\end{document}